\newcommand{\CLASSINPUTtoptextmargin}{2.2cm}

\documentclass[conference]{IEEEtran}

\IEEEoverridecommandlockouts

\usepackage{cite}
\usepackage{amsmath,amssymb,amsfonts}

\usepackage{graphicx}
\usepackage{textcomp}
\usepackage{xcolor}
\usepackage{balance}
\usepackage{algpseudocode}
\usepackage{algorithm}
\usepackage{amsmath}
\usepackage{lipsum}
\usepackage{fancyhdr}

\newcommand\blfootnote[1]{%
	\begingroup
	\renewcommand\thefootnote{}\footnote{#1}%
	\addtocounter{footnote}{-1}%
	\endgroup
}

\ifCLASSOPTIONcompsoc
\usepackage[caption=false,font=footnotesize,labelfont=sf,textfont=sf]{subfig} 
\else
\usepackage[caption=false,font=footnotesize]{subfig}
\fi

\def\BibTeX{{\rm B\kern-.05em{\sc i\kern-.025em b}\kern-.08em
		T\kern-.1667em\lower.7ex\hbox{E}\kern-.125emX}}
	
\makeatletter
\newcommand*\titleheader[1]{\gdef\@titleheader{#1}}
\AtBeginDocument{%
	\let\st@red@title\@title
	\def\@title{%
		\vskip-1em\bgroup\normalfont\large\centering\@titleheader\par\egroup
		\vskip0em\st@red@title}
}
\makeatother

	\title{Platoon Leader Selection, User Association and Resource Allocation on a C-V2X based highway: A Reinforcement Learning Approach\\ \vspace{-1ex}}
	\titleheader{Accepted by 2023 IEEE International Conference on Communications (ICC), \copyright2023 IEEE}
	
\begin{document}	

\author{\normalsize
	Mohammad Farzanullah and Tho Le-Ngoc\\
	Department of Electrical \& Computer Engineering, McGill University, Montr\'{e}al, QC, Canada
	
} 
	
	\maketitle
	
	\begin{abstract}
	We consider the problem of dynamic platoon leader selection, user association, channel assignment, and power allocation on a cellular vehicle-to-everything (C-V2X) based highway, where multiple vehicle-to-vehicle (V2V) and vehicle-to-infrastructure (V2I) links share the frequency resources. There are multiple roadside units (RSUs) on a highway, and vehicles can form platoons, which has been identified as an advanced use-case to increase road efficiency. The traditional optimization methods, requiring global channel information at a central controller, are not viable for high-mobility vehicular networks. To deal with this challenge, we propose a distributed multi-agent reinforcement learning (MARL) for resource allocation (RA). Each platoon leader, acting as an agent, can collaborate with other agents for joint sub-band selection and power allocation for its V2V links, and joint user association and power control for its V2I links. Moreover, each platoon can dynamically select the vehicle most suitable to be the platoon leader. We aim to maximize the V2V and V2I packet delivery probability in the desired latency using the deep Q-learning algorithm. Simulation results indicate that our proposed MARL outperforms the centralized hill-climbing algorithm, and platoon leader selection helps to improve both V2V and V2I performance. \blfootnote{This work was supported in part by the Natural Sciences and Engineering Research Council of Canada and in part by Huawei Technologies Canada.}
	\end{abstract}
	
	\begin{IEEEkeywords}
		 New radio, cellular vehicle-to-everything, reinforcement learning, resource allocation 
	\end{IEEEkeywords}
	
	\vspace{-2ex}

	\section{Introduction}
	Cellular vehicle-to-everything (C-V2X) is a vehicular standard that enables communication between vehicles and other entities on the road, such as pedestrians and infrastructure, to increase road safety and efficiency \cite{abou2019cellular}. The C-V2X system consists of communication between different entities, such as vehicle-to-vehicle (V2V), vehicle-to-infrastructure (V2I), vehicle-to-network (V2N), and vehicle-to-pedestrian (V2P) communications. The C-V2X is envisioned to support high throughput, ultra-reliable, and low latency communications.
	
	3GPP has provided the 5G new radio (NR) standards for vehicular communications concerning 
	advanced use-cases in TS 22.886 \cite{CV2X_22886}, and evaluation methodology in TR 37.885 \cite{CV2X_37885}. One component of infrastructure will be the roadside units (RSUs). RSU is a stationary wireless C-V2X device that can exchange messages with vehicles and other C-V2X entities. It uses the PC5 side-link interface to communicate with the vehicles and transmit information about road signs and traffic lights \cite{CV2X_22886}. It can also receive information from the vehicles to make a dynamic map of the surroundings and share it with other vehicles/pedestrians. Furthermore, we consider the use-case of platooning, where multiple vehicles form a train-like structure and travel closely together in a line. The platoon leader (PL) organizes communications between vehicles. Vehicle platooning has been identified as an advanced use-case in \cite{CV2X_22886} and has gained significant interest since it reduces fuel consumption and traffic congestion. The RSU and platoon will need to exchange a maximum of 1200 bytes in 500 ms for real-time traffic updates and 600 bytes in 10 ms for conditional automated driving \cite{CV2X_22886}. We grouped these two requirements for an aggregate of 624 bytes in 10 ms. Moreover, the PL and members need to exchange 50-1200 bytes in 10 ms for cooperative driving and up to 2000 bytes in 10 ms for collision avoidance \cite{CV2X_22886}. We aggregate these service requirements and keep the exchange of 1200-2800 bytes in 10 ms for our simulations. An intelligent resource allocation (RA) design is necessary for these stringent requirements. 
	
	The 5G new radio (NR) C-V2X supports two RA modes for sidelink PC5 communications: mode 1, the under-coverage mode, and mode 2, the out-of-coverage mode \cite{sehla2022resource}. In mode 1, the gNB allocates the communication resources to vehicles. Meanwhile, in mode 2, the vehicles autonomously select the resources. For mode 2, the current RA technique used in standards is the sensing-based  Semi-Persistent Scheduling (SPS) algorithm, which periodically selects random resources. However, the probability of resource selection collision can be high, and many works have considered either improving the SPS algorithm or alternate techniques to increase reliability. \cite{yi2020enhanced} proposes a novel sensing-based SPS algorithm in an urban scenario to reduce the collision probability. The authors in \cite{yoon2021stochastic} considered a highway scenario and suggested a stochastic reservation scheme for aperiodic traffic. For the platooning use-case, \cite{segata2021critical} shows that the SPS algorithm can not achieve the required performance.
	
	Due to the fast channel variations in vehicular networks, centralized optimization schemes that require global channel state information (CSI) will no longer be feasible. The high CSI overhead and the corresponding increase in latency make such methods impractical. To deal with this issue, distributed RA algorithms have been suggested in the literature, e.g., \cite{masmoudi2019survey, allouch2022survey}. Furthermore, traditional optimization techniques have limitations, requiring complete information about the environment and needing to be retrained for rapidly varying environments \cite{alwarafy2021deep}. Recently, distributed multi-agent reinforcement learning (MARL) has been proposed as an alternative approach to resolving such issues. The authors in \cite{liang2019spectrum} used Deep Q Networks (DQN) for joint channel assignment and power allocation to maximize the V2V delivery probability and V2N sum-rate in an urban setting, where the V2V links share the time-frequency resources with V2N links. Inspired by these results, \cite{vu2022multi} used double DQN for a platoon-based scenario for the same objectives. \cite{parvini2021aoi} uses the actor-critic method for mode selection, subchannel selection, and power control in an urban platoon scenario to increase the transmission probability of cooperative awareness messages. The authors in \cite{cao2021resource} used the Monte-Carlo algorithm to select resource blocks to reduce the packet collision probability in a platooning scenario.
	
	This paper considers a highway C-V2X system consisting of multiple platoons. We consider the periodic payload delivery from PLs to RSUs, termed V2I links. Furthermore, we consider the periodic transmission of messages from PLs to members, termed V2V links. We assume a limited spectrum is available for the V2V and V2I transmission, and they share the frequency resources for efficient spectrum usage. Given this system, this paper formulates a dynamic PL selection, user association, channel assignment, and power level control to maximize the packet delivery probability for both V2V and V2I links. Reliability is defined as the successful transmission of the packet within a time constraint $T$ \cite{liang2019spectrum,vu2022multi}.
	
	We utilize MARL in a distributed manner. The RL works on a trial-and-error strategy, and each agent slowly improves the action taken based on the feedback from the vehicular environment. We use the Deep Q-learning algorithm, which DeepMind developed for Atari video games \cite{mnih2015human}. Deep Q-Learning has been used for joint channel assignment and power allocation in C-V2X systems \cite{liang2019spectrum, vu2022multi}. However, we also use Deep Q-learning for user association and PL selection. As per our knowledge, dynamic PL selection has not been investigated in the literature. In our work, there are multiple collaborative agents for PL selection, V2V joint channel assignment and power allocation, and V2I joint user association and power allocation. The objective is to increase reliability for both V2V and V2I links. Simulation results indicate that the proposed MARL algorithm can outperform other benchmarks, such as the hill-climbing algorithm, which requires global CSI at the central controller. Moreover, the dynamic PL selection offers a gain in V2V and V2I reliability.
	
	\vspace{-0.5ex}
	\section{System Model and Problem Formulation}
	\vspace{-0ex}

\begin{figure} [h]
		\centering
		\includegraphics[scale=0.32]{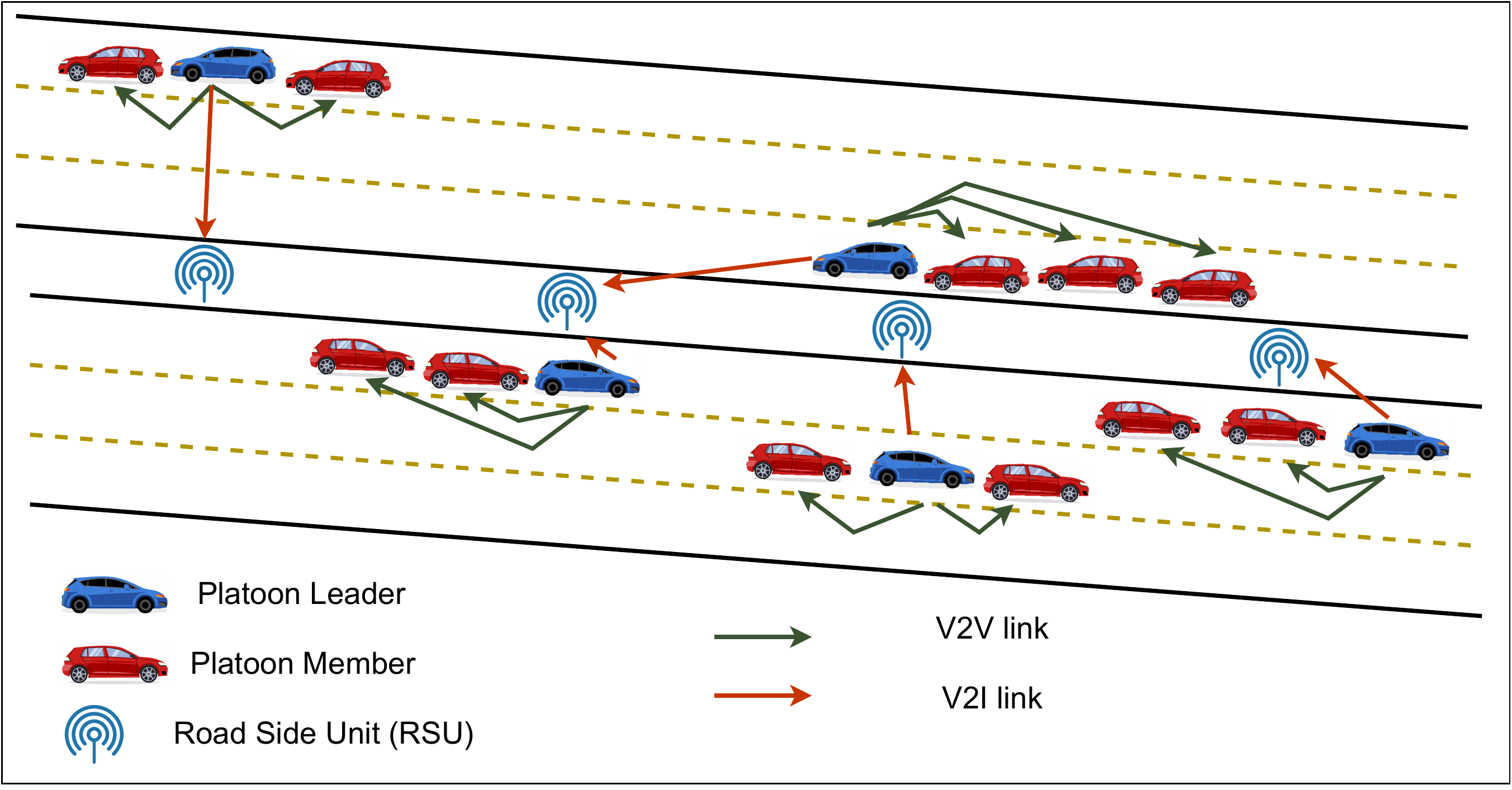}
		\vspace{-3ex}
		\caption{Illustrative C-V2X based highway, where multiple platoon leaders are transmitting to RSUs using the V2I links, and each platoon leader is transmitting to its platoon members using the V2V links. 
			\vspace{-4ex}
		} 
		\label{Fig:SystemModel} 
	\end{figure}

	As illustrated in Fig. \ref{Fig:SystemModel}, we consider a highway-based C-V2X System, outlined in \cite{CV2X_37885}. The highway consists of 3 lanes on both sides. The roadside units (RSUs) are placed in the middle of the highway, with 100 m between them. 
	We consider there are $K$ RSUs on the highway. Furthermore, we consider there are $M$ platoons, with $O$ vehicles in each platoon. The PL is required to share the real-time data with the RSU, referred to as V2I communications.
	
	Moreover, the PLs need to periodically transmit the cooperative awareness messages and the traffic data received from RSUs to the platoon members, referred to as V2V communications. 
	Each platoon is denoted as $m$, and the platoon leader is denoted as $m'$. In our simulations, the PL selection is dynamic, and all vehicles in a platoon are candidates for becoming the PL. The vehicles in a platoon are in the same lane, each with a single PL at a given time. It is considered that the vehicles are separated by a fixed distance of $d$ meters and are traveling with a velocity of $v$ m/s. It is considered that the vehicles and RSUs use a single antenna to transmit/receive the signal.
	
	We consider that a fixed and limited number of sub-bands are available for both V2V and V2I links, denoted as $N$. Each sub-band has a bandwidth of $W$. The PL needs to transmit a payload of size $B_{V2I}$ to the RSUs, and a payload of $B_{V2V}$ to the platoon members, within a time constraint of $T$. We assume that both V2V and V2I links use the same spectrum for efficient spectrum utilization. However, all the V2V and V2I links can interfere, making an intelligent design for interference management necessary. The set of RSUs, platoons, and sub-bands are denoted as $\mathcal{K}, \mathcal{M}$, and $\mathcal{N}$, respectively. Meanwhile, the set of members in each platoon is denoted as $\mathcal{V}_m, m=1,\dots, M$.
	In the paper, $L_{ab}$ refers to the large-scale fading power from transmitter $a$ to receiver $b$. The small-scale fading power from $a$ to $b$ in the sub-band $n$ is given by $g_{ab}[n]$. The $\rho_{ab}[n]$ is used as an indicator function set to 1 if the link $ab$ reuses the sub-band $n$ and 0 otherwise.
	
	
	The V2I links consist of communication from the PL to the RSU. Each PL $m'$ is required to transmit a fixed payload of $B_{V2I}$ to the RSUs within a time constraint of $T$. We assume each RSU has a sub-band preassigned to it. Each RSU $k$ will experience interference from other V2V and V2I links using the same sub-band $n$. Thus, the typical signal-to-interference (SINR) for a V2I link $m'k$ can be written as:
	
	\begin{align} \label{V2I_SINR} \small
		\text{SINR}_{m'k}[n] = \frac{P_{m'k}L_{m'k}g_{m'k}[n]}
		{I_k + \sigma_{k}^2}
	\end{align}

	Here, the interference at the RSU receiver is denoted by $I_k$	
	\begin{align} \label{V2I_Interference} \small
		I_k = \sum_{k \in \mathcal{K}} \sum_{a \neq m} \rho_{a'k}[n] P_{a'k} L_{a'k} g_{a'k}[n] + 
		\nonumber \\ \sum_{m=1}^{M} \sum_{o \in \mathcal{V}_{m}} \rho_{m'o}[n] P_{m'o} L_{m'o} g_{m'o}[n] 
	\end{align}	
	
	In (\ref{V2I_SINR}), the $\sigma_k^2$ is the noise power at the RSU. For simplicity, we assume that the noise power at all RSUs is equal.
	
	Given the SINR, the achievable rate for the V2I link from $m$ can be written as:
	
	\vspace{-2ex}
	\begin{align} \label{V2I:Rate} \small 
	R_m = W \log_2 (1 + \text{SINR}_{m'k}[n])
	\end{align}

	Meanwhile, the V2V link consists of communications between the PL and the members. Each PL $m'$ is required to transmit a fixed payload of size $B_{V2V}$ to each of its members $o$ in the time constraint $T$. The platoon member $o$ will experience interference from the other V2V and V2I links using the same sub-band $n$.
	The SINR for a platoon member $o$ in platoon $m$ can be written as:
	
	\begin{align} \label{V2V_SINR} \small
		\text{SINR}_{mo}[n] = \frac{P_{m'o}L_{m'o}g_{m'o}[n]}
		{I_o + \sigma_{o}^2}
	\end{align}

	Here, the interference at member $o$ is denoted by $I_o$	
	\begin{align} \label{V2V_Interference} \small
		I_o = \sum_{k \in \mathcal{K}} \sum_{m \in \mathcal{M}} \rho_{m'k}[n] P_{m'k} 	L_{m'o,k} g_{m'o,k}[n] + \nonumber \\
		\sum_{a=1, a \neq m}^{M}  \rho_{a'x}[n] P_{a'x} L_{a',mo} g_{a',mo}[n] 
	\end{align}	
	where $x$ denotes the platoon members in platoon $a$.
	
	Given the SINR, the achievable rate for the platoon member $o$ in platoon $m$ can be written as:
	
	\begin{align} \label{V2V:Rate} \small 
		R_{mo} = \frac{W}{O-1} \log_2 (1 + \text{SINR}_{mo}[n])
	\end{align}
	where we divide the bandwidth equally among the platoon members.

	\subsection{Problem Formulation}
	We consider a multi-objective optimization problem, where we simultaneously maximize the payload delivery probability for the V2V and V2I links. 
	For the V2I link, the objective for each PL is to transmit the payload $B_{V2I}$ to the RSUs within a time limit of $T$. This is given by 
	$\mathbb{P}(\Delta_T \sum_{t=1}^{T} \sum_{n=1}^{N} \rho_{mk}[n,t] R_m \geq B_{V2I}), \forall m \in \mathcal{M}$, where $\Delta_T$ is the channel coherence time.
	Meanwhile, for the V2V links, the objective is to maximize the delivery of payload $B_{V2V}$ within a time limit of $T$. This is given by 
	$\mathbb{P}(\Delta_T \sum_{t=1}^{T} \sum_{n=1}^{N} \rho_{mo}[n,t] R_{mo} \geq B_{V2V}), \forall  m \in \mathcal{M}, o \in \mathcal{V}_m$.
	
	Due to the spectrum sharing between the V2V and V2I links, we need to optimize two competing objectives of simultaneously maximizing the V2V and the V2I payload delivery probability. To achieve this, we use MARL for multiple objectives:
	\begin{enumerate}
		\item {Platoon Leader selection: For each platoon $m$, the PL will be selected dynamically and periodically. The platoon will decide which vehicle is the most suitable for being the leader so that both objectives can be met.}
		\item{Joint User Association and power allocation for V2I links: Each PL $m'$ will need to decide which RSU $k$ it needs to be served by, along with its transmit power level $P_{m'k}$}
		\item{Joint channel assignment and power allocation for V2V links: Each PL $m'$ needs to decide the channel $n$ and transmit power $P_{m'o}$ to transmit to its platoon members.}
	\end{enumerate}
	
	\section{Action Selection using Deep Reinforcement Learning}
	\subsection{Reinforcement Learning and Deep Q Learning}

	Reinforcement Learning (RL) is a discipline of Machine Learning (ML) where an agent can make a sequence of decisions by interacting with an environment. Based on the reward received by taking action, the agent learns to become intelligent. The agent aims to take actions that maximize the long-term cumulative reward. Markov Decision Process (MDP) is used to model an RL problem. 
	According to the Markov property, the current state captures all relevant information from history. 
	At each time-step $t$, the agent observes the environment through the state $s_t$ and takes an action $a_t$. The agent receives a reward $r_{t}$ and transitions into a new state $s_{t+1}$. 
	In RL, the goal of the agents is to maximize the cumulative reward $G_t$ it receives in the long run, given by 
	$		
	G_t = \sum_{k=0}^{\infty} \gamma^{k} r_{k+t} 
	$
	where $\gamma \in [0,1]$ represents the discount factor which reduces the present value of the future rewards.
	
		The action-value function $Q_{\pi} (s,a)$ is defined as the expected return by taking an action $a$ in state $s$ by following a policy $\pi$: 
		$
				Q_{\pi}(s,a) = \mathbb{E}_{\pi} [G_t | S_t = s, A_t = a] \nonumber
		$  
		where the expectation is taken over all possible transitions following the distribution $\pi$. The goal of RL is to find the optimal policy $\pi^{*}$ that maximizes the Q-function over all the policies.
	
	Q-learning is an off-policy RL algorithm that learns the value of an action $a$ in a state $s$. 
	It repetitively updates the action-value function for each state-action pair $(s,a)$ until they converge to the optimal action-value function $Q_{*}(s,a)$. The update equation is given by:
	$
			Q(s_t,a_t) \gets Q(s_t,a_t) + \alpha [r_{t+1} + \gamma \max_{a'} Q(s_{t+1},a') - Q(s_t,a_t)]
		$
	where $\alpha$ represents the learning rate. 
	If the Q-function is estimated accurately, the optimal policy $\pi^*$ at a given state $s$ would be to select the action $a^*$ that yields the highest value. 
	The Deep Q-Learning \cite{mnih2015human} uses a Deep Neural Network (DNN) as a function approximator to learn the Q-function. The state space is input to the DNN, and it learns to predict the Q-value for each output action. The state-action space is explored with a soft policy such as $\epsilon$-greedy,  
	where the agent takes random action at a given state $s_t$ at time $t$ with a probability of $\epsilon$. Otherwise, greedy action $a^* = \arg\max_{a \in \mathcal{A}} Q^* (s_t,a)$ is selected.
	The tuple $<s_t,a_t,r_{t},s_{t+1}>$ is stored in the replay memory at each time instance. At each time-step, a mini-batch is sampled from the replay memory to update the DNN parameters $\theta$,
		and the gradient descent algorithms are used to update the parameters $\theta$.

	\subsection{Multi-Agent Reinforcement Learning for optimization}
	
	In this section, we formulate the multi-agent RL algorithm for optimization. There will be three different types of agents, all based within the vehicles in the platoon. The first agent type will be the PL selection, which will dynamically decide the PL every 100 ms. The second type of agent will be the V2V platoon, which needs to determine the joint channel assignment and power allocation. Meanwhile, the third type of agent will be the V2I agent, which will need to optimize joint user association and power allocation. All the agents will interact with the environment and learn to take appropriate actions by trial and error. Furthermore, we use a common reward for all the agents to ensure collaboration. Moreover, each agent has a separate DQN and only uses its own experience to train the DNN.
	
	We develop two phases for the MARL problem: training and testing. During the training phase, each agent can access the common reward to train the DQN. Meanwhile, during the testing phase, each agent uses the trained DQN to select the action.
	
	\subsubsection{State and Action Space for Platoon Leader Selection}
	
	For the platoon leader selection agent, the state space $\mathcal{Z}_{pl} (t)$ consist of the measurements at time-step $t$. The state space consists of the following measurements: $i)$ The large-scale fading information between all members within a platoon $m$, i.e., $\{ L_{ab} \}_{(a,b) \in \mathcal{V}_m}$;  
	$ii)$ The large-scale fading information between all vehicles in platoon $m$ to all RSUs, i.e., $\{L_{ok}\}_{k \in \mathcal{K}, o \in \mathcal{V}_m}$. Meanwhile, the action space consists of the PL selection. The action is updated every 100 ms.
	
	\subsubsection{State and Action Space for V2V agent}
	The state space of the V2V agent, denoted by $\mathcal{Z}_{v2v} (t)$, consists of the measurements from the last time-step $t$ and consists of the following groups: $i)$ Direct channel measurements from the PL $m'$ to the members, i.e., $\{L_{m'o}g_{m'o}[n]\}_{o \in \mathcal{V}_m}$
	$ii)$ The interfering channels from other PLs sharing the same sub-band with the V2V agent $m$, which occupies the sub-band $n$, i.e., $\{\rho_{a'x}[n] P_{a'x} L_{a',mo} g_{a'_mo} \}_{a \in \mathcal{M}, a \neq m, x \in \mathcal{V}_a}$ 
	$iii)$ The interfering channels from the V2I links to the RSU, i.e., $\{\rho_{m'k}[n] P_{m'k} L_{m'o,k} g_{m'o_k}\}_{o \in \mathcal{V}_m , k \in \mathcal{K}}$ 
	$iv)$ The remaining payload and time limitation after the current time-step.
	
	Meanwhile, the action space consists of the combination of sub-band selection and power allocation. The sub-band consists of $N$ disjoint sub-bands, and the power levels are broken down into multiple discrete levels in the range $[0, P_d]$, where $P_d$ denotes the maximum power.
	
	\subsubsection{State and Action Space for V2I agent}
	The state space of the V2I agent, denoted by $\mathcal{Z}_{v2i} (t)$, for the PL $m'$, consists of the measurements from the last time-step $t$ and consists of the following groups:
	$i)$ Direct channel measurements from the PL $m'$ to all the RSUs, i.e., $\{L_{m'k} g_{m'k}\}_{k \in \mathcal{K}}$
	$ii)$ The remaining payload and time remaining after current time-step.
	$iii)$ The training iteration number $e$ and the agent's probability of random action selection $\epsilon$.
	
	The action space consists of the RSU selected and the power level. We assume that each RSU uses a fixed sub-band. There are $K$ RSUs to select from and transmit at a power divided into multiple discrete levels in the range $[0, P_d]$.
	
	\subsubsection{Reward function design}
	We use a common reward for all the agents in our proposed MARL design to ensure collaboration. We have a multi-objective problem, which is to maximize the payload delivery probability for the PL to RSU V2I links, and maximize the payload delivery probability for the PL to platoon members V2V links, within the time constraint $T$. The V2V and V2I agents need to select actions to minimize interference between each other.
	To achieve this purpose, we define the reward at time-step $t$, denoted as $r_t$, as:
	
	\vspace{-2ex}
	\begin{align} \label{reward} \small
		r_t = w_c \sum_{m=1}^{M} \sum_{o \in \mathcal{V}_m} U_{mo}(t) +
		w_d \sum_{m=1}^{M} V_m (t)
	\end{align}
	where $U_{mo}(t)$ is the contribution towards the reward of the V2V link $m'o$ and $V_m (t)$ is the contribution of the V2I link from PL $m'$ to RSU. Furthermore, $w_c , w_d \in [0,1]$ are weights to balance the two objectives.
	
	$U_{mo}(t)$ is the achievable rate of the PL to platoon member link $mo$, defined as:
	\begin{align} \label{reward:V2V} \small
	U_{mo}(t) = \left\{ \begin{array}{l}
		R_{mo}(t), \quad \text{if} \, B_{mo}(t) \ge 0, \\
		U, \,\,\,\quad\quad\quad\,\,  \text{otherwise}. 
	\end{array} \right.
	\end{align}
	where $B_{mo}(t)$ is the remaining payload for the V2V link $mo$ at time-step $t$. Furthermore, if the payload intended for link $mo$ has been delivered, the agent is given an award $U$, which needs to be greater than the maximum rate achievable, to indicate to the agent the successful transmission of the payload. $U$ is a hyperparameter that needs to be adjusted empirically \cite{liang2019spectrum}.
	
	Similarly, $V_m (t)$ is the achievable rate of the PL to RSU link, defined as:
	\begin{align} \label{reward:V2I} \small
	V_{m}(t) = \left\{ \begin{array}{l}
		R_{m}(t), \quad \text{if} \, B_{m}(t) \ge 0, \\
		V, \,\,\,\quad\quad\quad\,\,  \text{otherwise}. 
	\end{array} \right.
	\end{align}	
	where $B_m(t)$ is the payload that PL $m'$ needs to transmit to the RSUs. $V$ is a hyperparameter, which needs to be greater than the maximum achievable rate of the V2I link.
		
	\subsection{Training Algorithm and Testing Strategy}
	
	We devise the problem as an episodic setting, where each episode corresponds to the time limit $T$ for the V2V and V2I links to complete their transmission. Each episode consists of multiple time-steps $t$. The vehicle location and large-scale fading are updated every 100 ms \cite{CV2X_37885}. Meanwhile, the small-scale fading is updated at each time-step $t$, changing the state space for the V2V and V2I agents and prompting the agents to adjust their actions. Each agent stops its transmission once its payload has been delivered. 
	The training is centralized, where each agent has access to the common reward $r_t$. Deep Q-Learning is used to train the agent. 
	The algorithm is outlined in Algorithm 1.
\setlength{\textfloatsep}{0pt}
		\begin{algorithm}[htb!]   \small

			\caption{\small Training Algorithm}	
			\begin{algorithmic}[1] 
				\State Initiate the environment and generate the V2I and V2V links
				\State Initiate the DQN with random parameters $\theta$
				\For{each episode i}
				\If{i\%20 = 0}
				\State Update the vehicle locations and large-scale fading
				\For{each PL selection agent $n_1$}
				\State Observe the state $s_{t1}$ for PL selection and take action $a_{t1}$
				\EndFor
				\EndIf
				\For{each time-step $t$}
				\State Update small-scale channel fading
				\For{each V2V agent $n_2$}
				\State Observe the state $s_{t2}$				
				\EndFor
				\For{each V2I agent $n_3$}
				\State Observe the state $s_{t3}$				
				\EndFor
				\State All agents take actions simultaneously according to the $\epsilon-$greedy policy and receive the common reward $r_t$ 
				
				\For{each agent \{$n_1$, $n_2$, $n_3$\}}
				\State Observe the next state $s_{t+1}$		
				\State Store $e_t=[s_t, a_t, r_{t}, s_{t+1}]$ in the replay memory				
				\EndFor
				\EndFor
				\For{each agent \{$n_1$, $n_2$, $n_3$\}}
				\State Uniformly sample mini-batch data $\mathcal{D}$ from replay memory
				\State Train the deep Q-networks using the mini-batch data.			
				\EndFor		
				\EndFor 
				
			\end{algorithmic} 
	
\end{algorithm} 

	During the testing phase, each agent observes the state. The state is input to the trained DQN, which is used to select the action. The testing is implemented in a distributed manner, where each agent takes action based on their local state observation only.

	\section{Illustrative Results}
	This section presents the simulation results to illustrate the performance of our algorithm. We consider a highway setting as described in TR 37.885 \cite{CV2X_37885}, with the carrier frequency of 6 GHz. The technical report provides all details, such as evaluation scenarios, vehicle drop and mobility modeling, RSU deployment, and channel models for V2V and V2I links. The small-scale fading is modeled as Rayleigh fading. We consider a highway with a length of 1 km, with 3 lanes for traffic on both sides. The RSUs are placed in the middle of the highway, with a distance of 100 m between them. We use option A in UE drop options in Section 6.1.2 of TR 37.885 \cite{CV2X_37885}. The type 3 vehicles (bus/tracks) are used, with a length of 13 m and 2 m distance between each in a platoon. All vehicles travel with a velocity of 140 km/h.  Each V2V platoon consists of 3 vehicles. Moreover, the antenna on vehicles is placed in the middle of each vehicle. As per TS 22.185 \cite{CV2X_22886}, the V2V and V2I links need to complete their transmission in 10 ms. However, we set $T$ as 5 ms, assuming the other 5 ms will be used for communication in other directions, i.e., platoon member to platoon leader. The main simulation parameters are listed in Table \ref{Table:Parameters}.
	
	\begin{table} [htb!]
	\caption{Simulation Parameters}	
	\centering %
	\begin{tabular}{|c | c|}				
		\hline	
		Carrier frequency & $6$ GHz \\ \hline	
		Bandwidth of each sub-band & $1$ MHz \\ \hline 	
		Number of sub-bands $N$ & 2 \\ \hline
		Number of RSUs $K$ & 11 \\ \hline
		Number of platoons $M$ & [4,6] \\ \hline
		Number of vehicles in each platoon $O$ & 3 \\ \hline
		Vehicle velocity $v$ & 140 km/h \\ \hline
		Tx power for V2V links & [23, 15, 5, -100] dBm \\ \hline
		Tx power for V2I links & [23, -100] dBm \\ \hline
		Vehicle Antenna gain & 3 dBi \\ \hline
		Vehicle receiver noise figure & 9 dB \\ \hline
		Noise PSD & -169 dBm/Hz \\ \hline
		Time constraint $T$ & 5 ms \\ \hline
		Platoon Leader update interval & 100 ms \\ \hline
		V2V payload $B_{V2V}$ & [1200,.....,2800] bytes \\ \hline
		V2I payload $B_{V2I}$ & 624 bytes \\ \hline
	\end{tabular} \vspace{0ex}
	\label{Table:Parameters}
	\end{table}	

	The DQN was implemented in Python using the Tensorflow package. The DNN for all 3 types of agents consisted of 3 hidden layers. The DNN of PL selection agents had 71, 35, and 17 neurons, the DNN of V2V agents had 100, 50, and 24 neurons; and the DNN of V2I agents had 166, 83, and 40 neurons in their hidden layers, respectively. The rectified linear unit (ReLU) was us as the activation function for all 3 types of agents. RMSProp was used for optimization for all agents, and learning rates of 0.0001, 0.0001, and 0.001 were used for PL selection agents, V2V agents, and V2I agents, respectively. The $w_c$ and $w_d$ in (\ref{reward}) were set as 0.3 and 0.7, respectively. Meanwhile, the hyperparameters $U$ in (\ref{reward:V2V}) and $V$ in (\ref{reward:V2I}) were selected as 25 and 15, respectively. The training phase consisted of 2000 episodes, and the testing was performed for 100 episodes. The $\epsilon$-greedy policy was used during the training, and the value of $\epsilon$ was reduced linearly from 1 to 0.02 for 1600 episodes. The training was performed setting $B_{V2V}$ as 2400 bytes and was varied between 1200-2800 bytes during testing. Meanwhile, $B_{V2I}$ was set to 624 bytes during the training and testing phases.
	
	We developed three benchmarks for comparison: $1)$ Hill-climbing algorithm \cite{russell2010artificial}: Hill-climbing algorithm is a local search optimization method, guaranteed to reach a local optimum. It is a centralized iterative algorithm, which starts with a random solution, and then iteratively keeps improving it until it reaches an optimum. The algorithm is used as an upper benchmark in our paper. $2)$ Greedy Algorithm: Each agent uses the best link available to transmit at maximum power. $3)$ RL algorithm without PL selection: We run the RL algorithm, fixing the leading vehicle in each platoon as the platoon leader. This is to show the effectiveness of PL selection agents.

	\begin{figure}[htb!]
	\centering
	\vspace{-3ex}
	\includegraphics[scale=0.55]{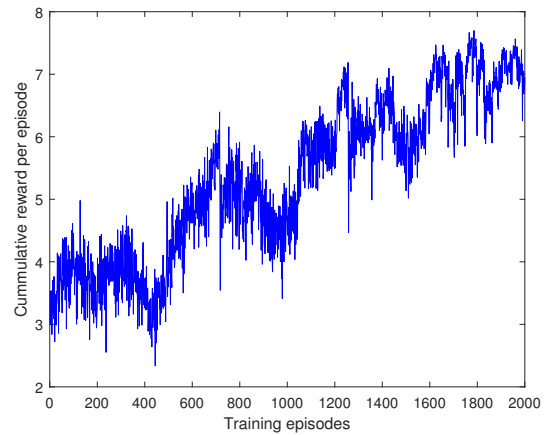}
	\vspace{-2ex}
	\caption{Cumulative reward per episode for $M$ = 4
	} \vspace{-1ex}
	\label{Fig:Reward_4Platoons} 
	\end{figure} 

	Fig. \ref{Fig:Reward_4Platoons} shows the cumulative reward for each episode for the case of 4 platoons. The reward increases during training, indicating that the agents can collaborate.

%
%
%
	\begin{figure}[htb!] 
		\centering
		\subfloat[Average V2V payload delivery probability]{\includegraphics[scale=0.64]{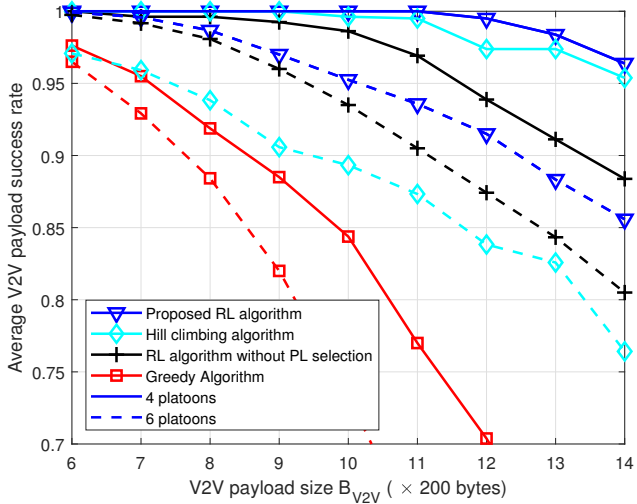}
			\label{Fig:combined_V2V}}
		\hfil 
		\vspace{-0ex}
		\subfloat[Average V2I payload delivery probability]{\includegraphics[scale=0.64]{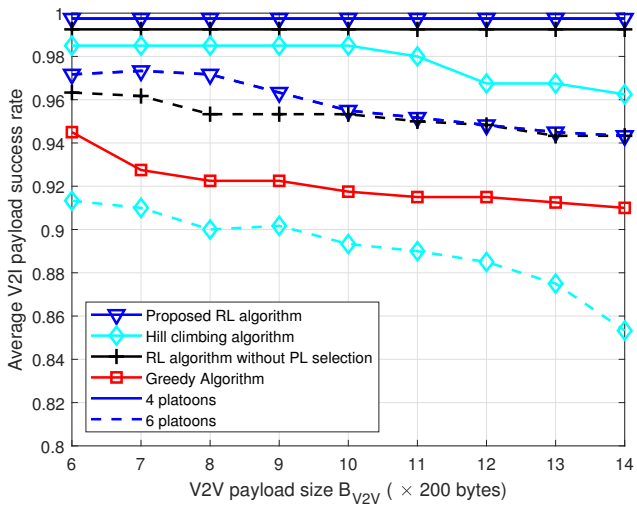}
			\label{Fig:combined_V2I}}

		\caption{V2V and V2I performance for $M = 4$ and $M=6$.}
		\label{Fig:combined}
	\end{figure}
	
	Fig. \ref{Fig:combined} shows the reliability of V2V and V2I links as we increase $B_{V2V}$ for the cases of 4 and 6 platoons. Fig. \ref{Fig:combined_V2V} shows that the V2V performance decreases as we increase the packet size. This is because a larger payload size requires a longer time to transmit. For 4 platoons, we achieve a reliability of 1 for up to 2200 bytes, outperforming all the other benchmarks. Fig. \ref{Fig:combined_V2I} shows that for 4 platoons, the payload success rate for V2I links is 0.9975 for all cases. However, the hill-climbing and greedy algorithm performance decrease, indicating more significant interference as we increase V2V payload size. When we increase the number of platoons to 6, the performance gap between the proposed algorithm and the hill-climbing algorithm increases, which shows the superiority of our algorithm for a higher number of agents. Furthermore, it can be seen that dynamic platoon leader selection improves performance for both V2V and V2I links.
	
%

	\section{Conclusion}
	
	In this work, we proposed a distributed multi-agent reinforcement learning algorithm to optimize the performance of the V2V and the V2I links. The V2V and V2I links used the same spectrum, making an intelligent resource allocation design necessary to manage interference. Each platoon leader had an agent for joint channel assignment and power allocation for V2V links, and another agent for joint user association and power allocation for V2I links. Further, another agent was able to select the platoon leader, to maximize the reliability of both V2V and V2I links. Based on RL, the agents could collaborate to take suitable actions. The proposed approach is decentralized, and the agents were able to make decisions based on their local state observations only. Simulation results indicate that the proposed algorithm could perform well for variable V2V packet size and different numbers of platoons, outperforming the centralized hill-climbing algorithm. Moreover, the PL selection improved reliability for both V2V and V2I links. In the future, we plan to investigate if our approach generalizes well for different channel conditions and topologies.

	\balance
	\bibliographystyle{IEEEtran}
	\bibliography{PLRL}
	\balance \balance
	
	
\end{document}